\documentclass[conference]{IEEEtran}
\usepackage{graphicx}
\usepackage{color}
\usepackage[usenames,dvipsnames,svgnames,table]{xcolor}
\DeclareGraphicsExtensions{.eps}

\usepackage{algorithm}
\usepackage{algpseudocode}
\usepackage[english]{babel}
\usepackage{blindtext}

\usepackage{siunitx}
\usepackage{cite}
\ifCLASSINFOpdf
\else
\fi
\usepackage[cmex10]{amsmath}
\usepackage{footnote}
\makesavenoteenv{tabular}
\makesavenoteenv{table}

\usepackage[footnotesize, tight]{subfigure}

\IEEEspecialpapernotice{(Invited Paper)}

\usepackage[top=0.7in, bottom=1.1in, left=0.625in, right=0.625in]{geometry}

\DeclareRobustCommand*{\IEEEauthorrefmark}[1]{%
  \raisebox{0pt}[0pt][0pt]{\textsuperscript{\footnotesize\ensuremath{#1}}}}

\begin{document}
\title{Dynamic Double Directional Propagation Channel Measurements at 28 GHz }


%

%
\author{
    \IEEEauthorblockN{C. U. Bas\IEEEauthorrefmark{1}, {\it Student Member, IEEE},
    R. Wang\IEEEauthorrefmark{1}, {\it Student Member, IEEE},
    S. Sangodoyin\IEEEauthorrefmark{1}, {\it Student Member, IEEE}, \\
    S. Hur\IEEEauthorrefmark{3}, {\it Member, IEEE},
    K. Whang\IEEEauthorrefmark{3}, {\it Member, IEEE},
    J. Park\IEEEauthorrefmark{3}, {\it Member, IEEE}, \\
    J. Zhang\IEEEauthorrefmark{2}, {\it Fellow, IEEE}, 
    A. F. Molisch\IEEEauthorrefmark{1}, {\it Fellow, IEEE}  
    }\\
    \IEEEauthorblockA{\IEEEauthorrefmark{1}University of Southern California, Los Angeles, CA, USA,}
    
    \IEEEauthorblockA{\IEEEauthorrefmark{2}Samsung Research America, Richardson, TX, USA}
    
     \IEEEauthorblockA{\IEEEauthorrefmark{3}Samsung Electronics, Suwon, Korea}
}


\maketitle
\IEEEpeerreviewmaketitle

\begin{abstract}
This paper presents results from the (to our knowledge) first {\em dynamic} double-directionally resolved measurement campaign at mm-wave frequencies for an outdoor microcellular scenario. The measurements are performed with USC's real-time channel sounder equipped with phased array antennas that can steer beams electrically in microseconds, allowing directional measurements in dynamic environments. Exploiting the phase coherency of the setup, the multi-path components can be tracked over time to investigate the temporal dependencies of the channel characteristics. We present results for time-varying path-loss, delay spread, mean angles and angular spreads observed at the transmitter (TX) and receiver (RX) in the presence of moving vehicles and pedestrians. Additionally, we investigate excess losses observed due to blockage by vehicles and compare the cases when TX and RX are using fixed beams or when they are capable of adjusting beam directions dynamically.  
\end{abstract}

\section{Introduction} \label{sec_intro}

CISCO estimated the global mobile data traffic in 2016 as \SI{7} exabytes (EB, $10^{18}$ bytes) per month and forecast it to reach \SI{49}{EB} per month by 2021 \cite{forecast2017cisco}. One commonly anticipated way to meet this ever-growing demand is to utilize the idle spectrum available at millimeter-wave (mm-wave) frequencies. Especially, \SI{28}{GHz} band attracts a lot of interest thanks to comparatively lower hardware and implementation costs due to relatively lower carrier frequency.

An accurate propagation channel model is essential for designing and validating a wireless communication system. Due to shorter wavelengths at mm-wave frequencies, the nature of the propagation channels in that regime can be significantly different from those sub \SI{6}{GHz} bands. Consequently, there has been a growing interest in wireless propagation channel measurements at mm-wave frequencies \cite{Molisch_2016_eucap}. It is commonly accepted that mm-wave systems will have to use beam-forming arrays to overcome the higher pathloss experienced at higher carrier frequencies \cite{Roh2014millimeter}\cite{Molisch_2016_eucap}. Hence an accurate model for the angular statistics is crucial to assess system performance. Moreover, in case of dynamic channels, an accurate model of temporal characteristics of these angular statistics is imperative for system design.

In this work, we use a \SI{28}{GHz} real-time, double-directional mm-wave channel sounder allowing us to perform directional measurements in dynamic channels \cite{bas_realjournal_2017, bas_2017_realtime}. Instead of a rotating horn, our channel sounder uses phased array antennas to form beams that can be steered electronically. By using measurements with this channel sounder, we investigate time-varying PDP, path-loss, delay spread, mean angles and angular spreads observed at the transmitter (TX) and receiver (RX) in the presence of moving vehicles and pedestrians. Additionally, we investigate excess losses observed due to blockage by vehicles and compare the cases when TX and RX are using fixed beams or when they are capable of adjusting directions of beams dynamically. To the best of our knowledge, no such measurements exist in the literature up to now. Most of the directional measurements of sounders used in mm-wave channels are based on  rotating horn antennas which are not suitable for the intended measurements here \cite{MacCartney_2017_flexible,hur_synchronous_2014,Haneda_2016_omni}.

Conversely, most of the papers in the literature investigating the dynamic channel characteristics for the mm-wave focus on the shadowing effects of moving vehicles or humans  {\em when the TX and RX antennas are fixed} (they may be directional, but cannot {\em adapt} their directions). In \cite{Marinier1998temporal}, the authors performed \SI{30}{GHz} indoor measurements with omnidirectional antennas to investigate the human-induced variations. \cite{maccartney2017rapid} proposes a model for the fading due to human blockage at \SI{73.5}{GHz} with Markov models by using measurements performed with directional horn antennas with different beam-widths. \cite{collonge2004influence} measured the impact of people walking through/near the significant paths between TX and RX at \SI{60}{GHx}. They found that strongest outages (additional attenuation 15 dB or more) occurred when both TX and RX are at/below height of the walking people but smaller when one of the antennas was 2m or higher.\cite{Park_2017_vehicular} presents the blockage characteristics at \SI{28}{GHz} due to a stationary car. They compare path loss, delay spread and angular spread with and without vehicular blockage for static measurements. The references \cite{Weiler_et_al_2016_quasideterministic} and \cite{Weiler_et_al_2016_WCL} investigate shadowing characteristics due to vehicles and humans with omnidirectional antennas at \SI{28}{GHz}. Both \cite{Semkin_et_al_2015_EuCAP} at \SI{76}{GHz} and \cite{Sato_et_al_2001_V2I_RoF} at \SI{36}{GHz} observed \SI{20}{dB} attenuation due to blockage by a truck.

However, none of those previous works is capable of providing directional information in dynamic channels. Hence they can't provide insights on the temporal dependencies of the angular statistics, which are crucial for a mobile system operating with beam-forming arrays.

The rest of the paper is organized as follows. Section \ref{sec_design} introduces the channel sounder setup and the configuration used in the measurements. Section \ref{sec_meas} describes the measurement environment and scenarios. Section \ref{sec_eval} discusses the evaluation performed for the measurements results discussed in Section \ref{sec_result}. Finally, Section \ref{sec_conc} summarizes results and suggests directions for future work.

\section{Channel Sounder Setup} \label{sec_design}

  \begin{table}[tbp]\centering
\small
  \caption{Sounder Parameters}
  \renewcommand{\arraystretch}{1.2}
\begin{tabular}{l|c}
    \hline
    \multicolumn{2}{c}{\textbf{Hardware Specifications}} \\ \hline \hline
    Center Frequency & 27.85 GHz\\
    Instantaneous Bandwidth & 400 MHz\\
    Antenna array size & 8 by 2 (for both TX and RX) \\
    Horizontal beam steering & -45 to 45 degree \\
    Horizontal 3dB beam width & 12 degrees\\
    Vertical 3dB beam width & 22 degrees\\
    Horizontal steering steps & 10 degrees\\
    Beam switching speed & 2$\mu s$ \\
    TX EIRP & 36 dBm (max 57 dBm) \\
    RX noise figure & $\le$ 5 dB \\ 
    ADC/AWG resolution & 10/15-bit \\
    Data streaming speed & 700 MBps \\ \hline
    \multicolumn{2}{c}{\textbf{Sounding Waveform Specifications}} \\ \hline \hline
    Waveform duration & 2 $\mu s$ \\
    Repetition per beam pair & 1 \\
    Number of tones & 801 \\
    Tone spacing & 500 kHz \\
    PAPR & 0.4 dB \\ 
    Total sweep time & 400 $\mu s$ \\ 
    MIMO repetition per burst & 20   \\ 
    Burst repetition & 16.66 Hz \\ \hline 
  \end{tabular} \label{specs}
\end{table}

In this campaign, we used a switched-beam, wide-band mm-wave sounder with 400 MHz real-time bandwidth \cite{bas_2017_realtime}.The sounding signal is a multi-tone signal which consists of equally spaced 801 tones covering 400 MHz. A low peak to average power ratio (PAPR) of $0.4$ dB is achieved by manipulating the phases of individual tones as suggested in \cite{Friese1997multitone}. This allows us to transmit with power as close as possible to the 1 dB compression point of the power amplifiers without driving them into saturation.  

Both the TX and the RX have phased arrays capable of forming beams which can be electronically steered with $5^{\circ}$ resolution in the range of $[-45^{\circ}, 45^{\circ}]$ in azimuth and $[-30^{\circ}, 30^{\circ}]$ in elevation. Compared to rotating horns, this decreases the measurement time for each RX location from tens of minutes to milliseconds. During this measurement campaign we only utilize a single elevation angle $0^{\circ}$ with 10 azimuth angles both for the TX and the RX. The total sweep time is \SI{400}{\mu s} for 100 total beam pairs. Moreover, thanks to the beam-forming gain, the maximum TX EIRP is 57 dBm, and the measurable path loss is 159 dB without considering any averaging or processing gain. By using GPS-disciplined Rubidium frequency references, we were able to achieve both short-time and long-time phase stability. Combined with the short measurement time this limits the phase drift between TX and RX, enabling phase-coherent sounding of all beam pairs even when TX and RX are physically separated and have no cabled connection for synchronization. Consequently, the directional power delay profiles (PDP) can be combined easily to acquire the omnidirectional PDP. Table \ref{specs} summarizes the detailed specification of the sounder and the sounding waveform. References \cite{bas_2017_realtime}, \cite{bas_2017_microcell} and \cite{wang_2017_stationarity} discuss further details of the sounder, validation measurements and prior channel sounding campaigns performed with the same channel sounder.

\section{Measurement Campaign} \label{sec_meas}

\begin{figure}[tbp]
        \centering\includegraphics[width=1\linewidth]{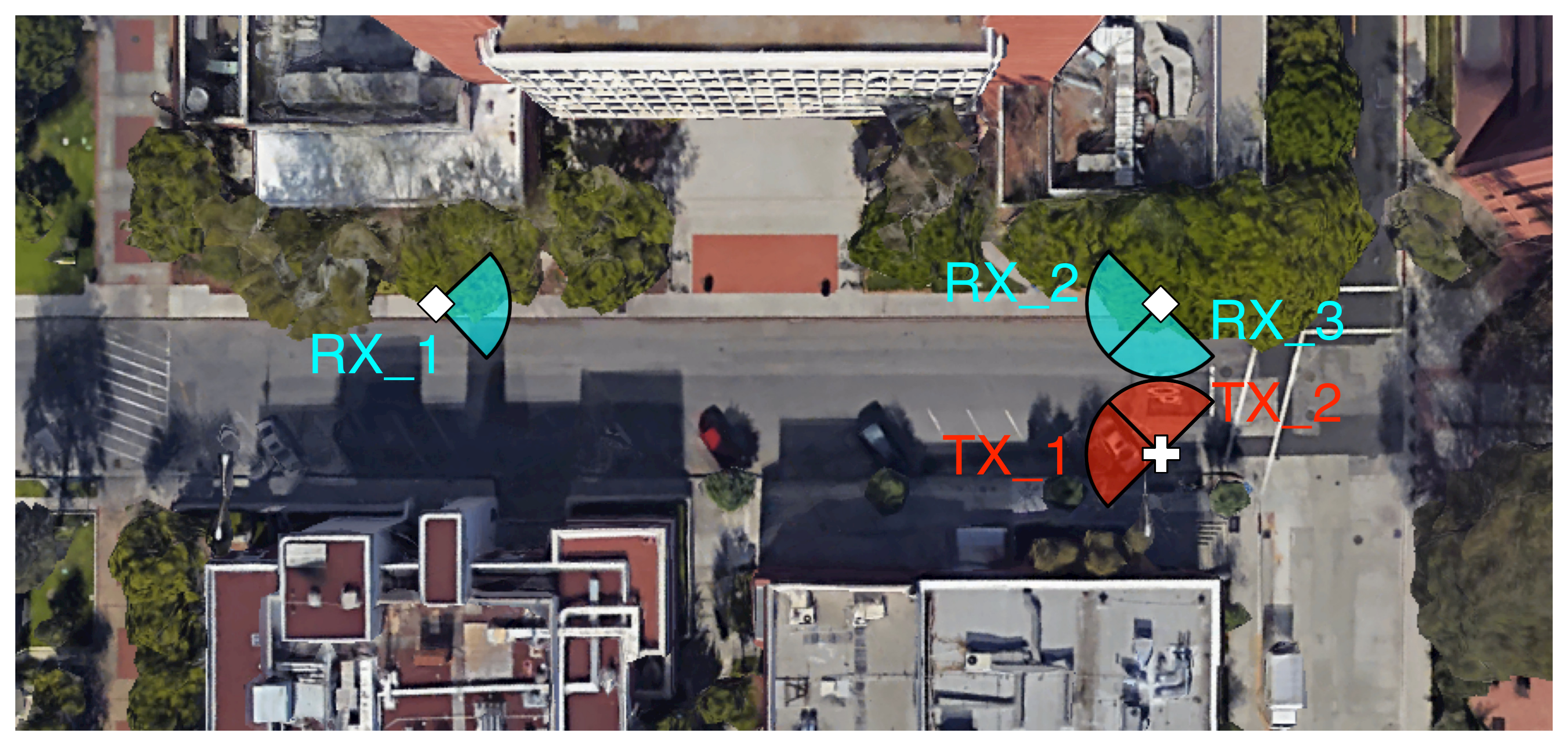}\caption{Measurement environment}\label{fig:loc_all}
\end{figure}

The measurements were performed on an urban street in the University of Southern California University Park Campus in Los Angeles, CA, USA. TX and RX were always placed across the street from each other as shown in Figure \ref{fig:loc_all}. The RX height is \SI{1.8}{m} while the TX height was changed to \SI{2.5}{m}, \SI{3.5}{m} and \SI{4.5}{m}. 

From the TX and RX orientations shown in the Figure \ref{fig:loc_all}, the following TX-RX pairs were measured:
\begin{itemize}
  \item Case 1: TX\_1 to RX\_1
  \item Case 2: TX\_1\ to RX\_2
  \item Case 3: TX\_2 to RX\_3
\end{itemize}

In Cases 1 and 3, a LOS exists when the environment is idle but it may or may not be blocked by passing cars, buses or trucks. For Case 2, although TX and RX have a visual LOS, they are placed so that TX and RX are out of each others visible azimuth range, i.e., their beams do not point towards each other.  

Each SISO measurement takes \SI{4}{\mu s} consisting of \SI{2}{\mu s} sounding waveform and \SI{2}{\mu s} guard time for electronical beam-switching. Although the channel sounder is capable of beam-steering with $5^\circ$ steps covering $\pm 45^\circ$, to decrease the measurement time we used every other beam resulting in $10^\circ$ angular resolution. Consequently, both TX and RX perform $90^\circ$ azimuth sweeps measuring 100 beam pairs in only $400 \mu s$. A single sweep of all possible combinations of TX and RX beams is called a MIMO snapshot. In a burst, 20 MIMO snapshots were measured without any idle time in between. This allows us to estimate Doppler shifts up to $\SI{\pm 1.25}{kHz}$ which corresponds to a maximum relative speed of \SI{48}{kph} at \SI{28}{GHz}, which is larger than the maximum permissible speed on the measured street.. Furthermore, these bursts of MIMO measurements were repeated 200 times with a period of \SI{60}{ms} to track evolution of channel parameters as they change due to moving objects in the environment. This configuration provides us a unique capability of performing double-directional measurements under dynamic conditions and investigate the effects of moving objects on the angular channel characteristics and Doppler spectrum.

\section{Data Evaluations} \label{sec_eval}
The directional power delay profile (PDP) for the TX beam and RX beam with the azimuth angles  $\theta_{TX}$  and  $\theta_{RX}$  is estimated as;
\begin{equation}
  \resizebox{.9 \linewidth}{!} 
{
$   P(\theta_{TX},\theta_{RX},\tau) = \bigg\vert \mathcal{F}^{-1} \left\{ H_{\theta_{TX},\theta_{RX}}\left(\vec{f} \right) ./ H_{cal}\left(\vec{f} \right) \right\} \bigg\vert ^2$
}
\end{equation}
where $\theta_{RX}\in[-45,45]$,  $\theta_{TX}\in[-45,45]$, $\mathcal{F}^{-1}$ denotes inverse Fourier transform, $H_{\theta_{TX},\theta_{RX}}(\vec{f})$ and $H_{cal}(\vec{f})$ are the frequency responses for TX beam $\theta_{TX}$ and RX beam $\theta_{RX}$ and, the calibration response respectively; $\vec{f}$ is the vector of the used frequency tones, and $./$ is element-wise division. The specular multi-path components (MPCs) were then detected by performing a 3D (DoA, DoD, and delay) peak search. The ghost MPCs due to sidelobes of the beams were filtered out as described in \cite{bas_realjournal_2017}

\section{Measurement Results} \label{sec_result}

\begin{figure}[tbp]
        \centering\includegraphics[width=1\linewidth]{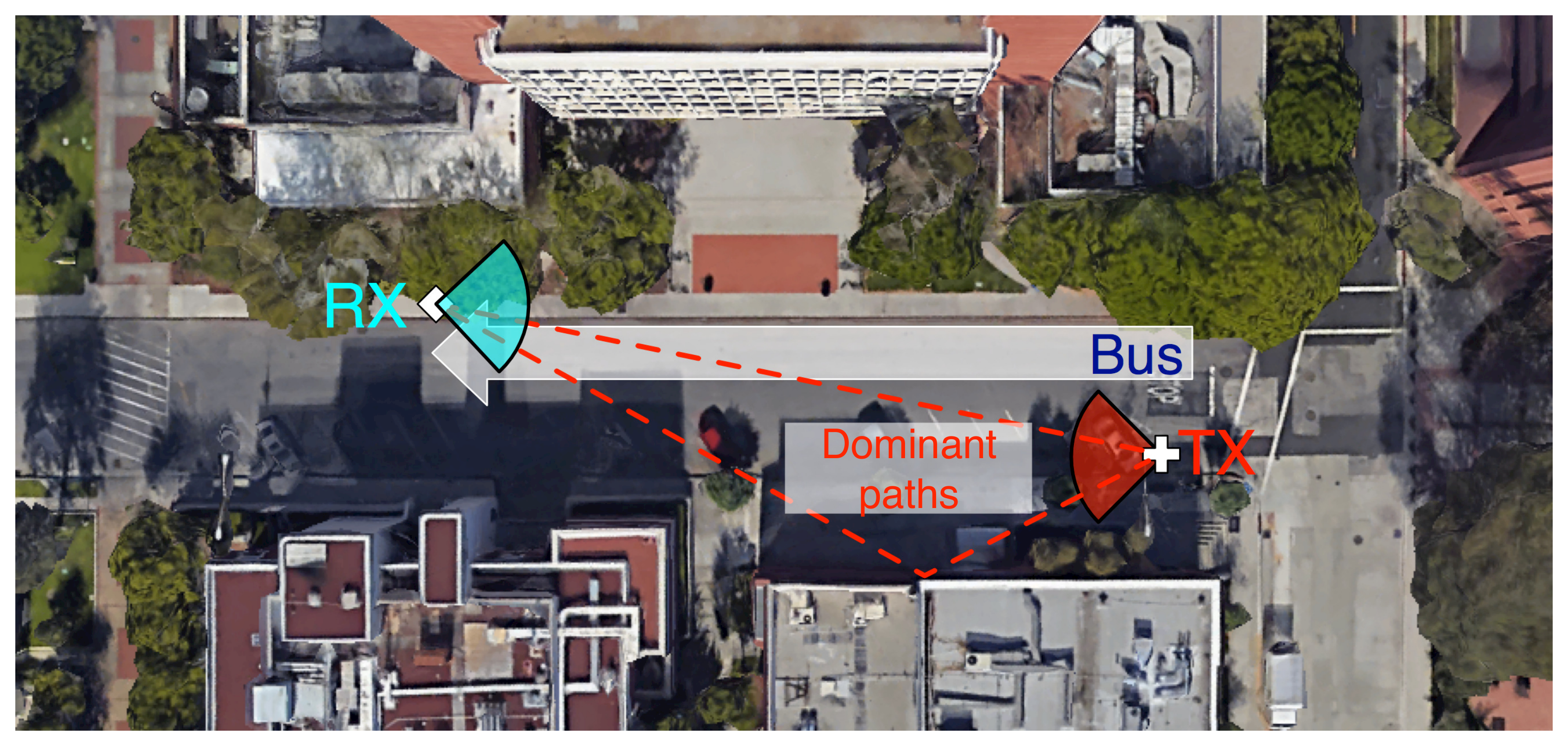}\caption{Blocked LOS}\label{fig:dynamic_1}
\end{figure}

\begin{figure}[tbp]
        \centering\includegraphics[width=1\linewidth]{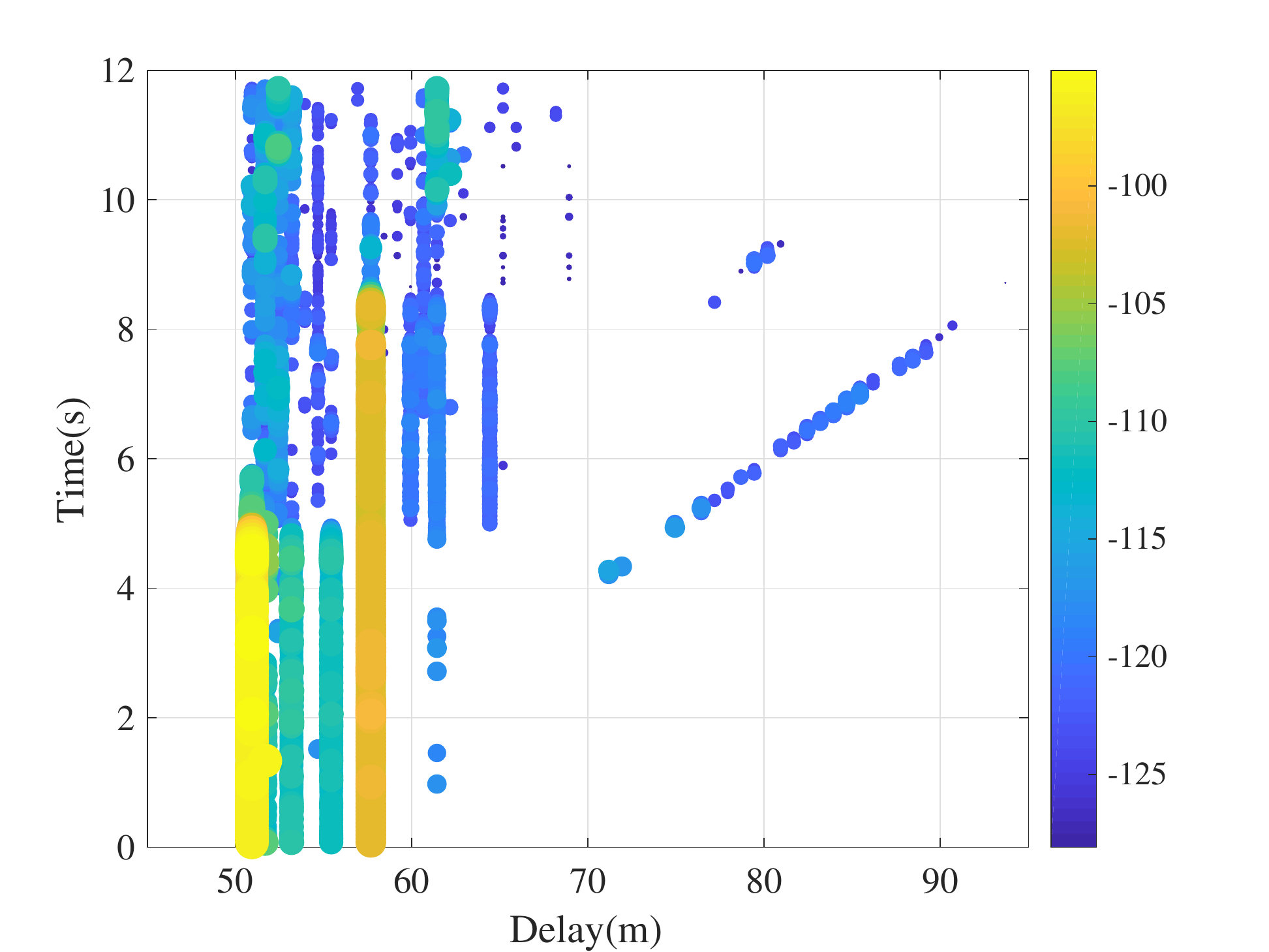}\caption{PDP vs time}\label{fig:paths_1}
\end{figure}

\begin{figure}[tbp]
        \centering\includegraphics[width=1\linewidth]{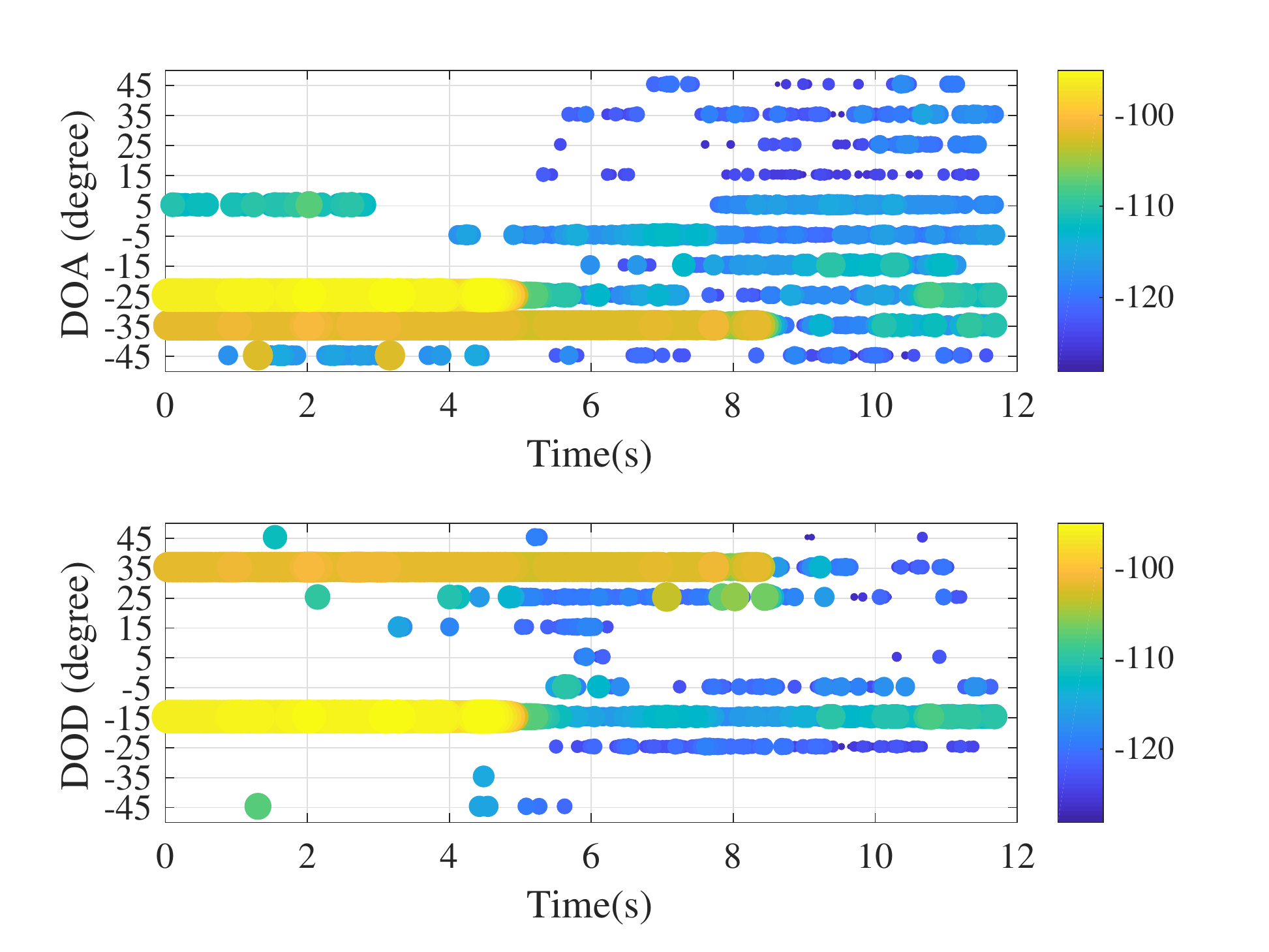}\caption{Power of the MPCs vs DOA/DOD and time}\label{fig:angles_1}
\end{figure}

\begin{figure}[tbp]
        \centering\includegraphics[width=1\linewidth]{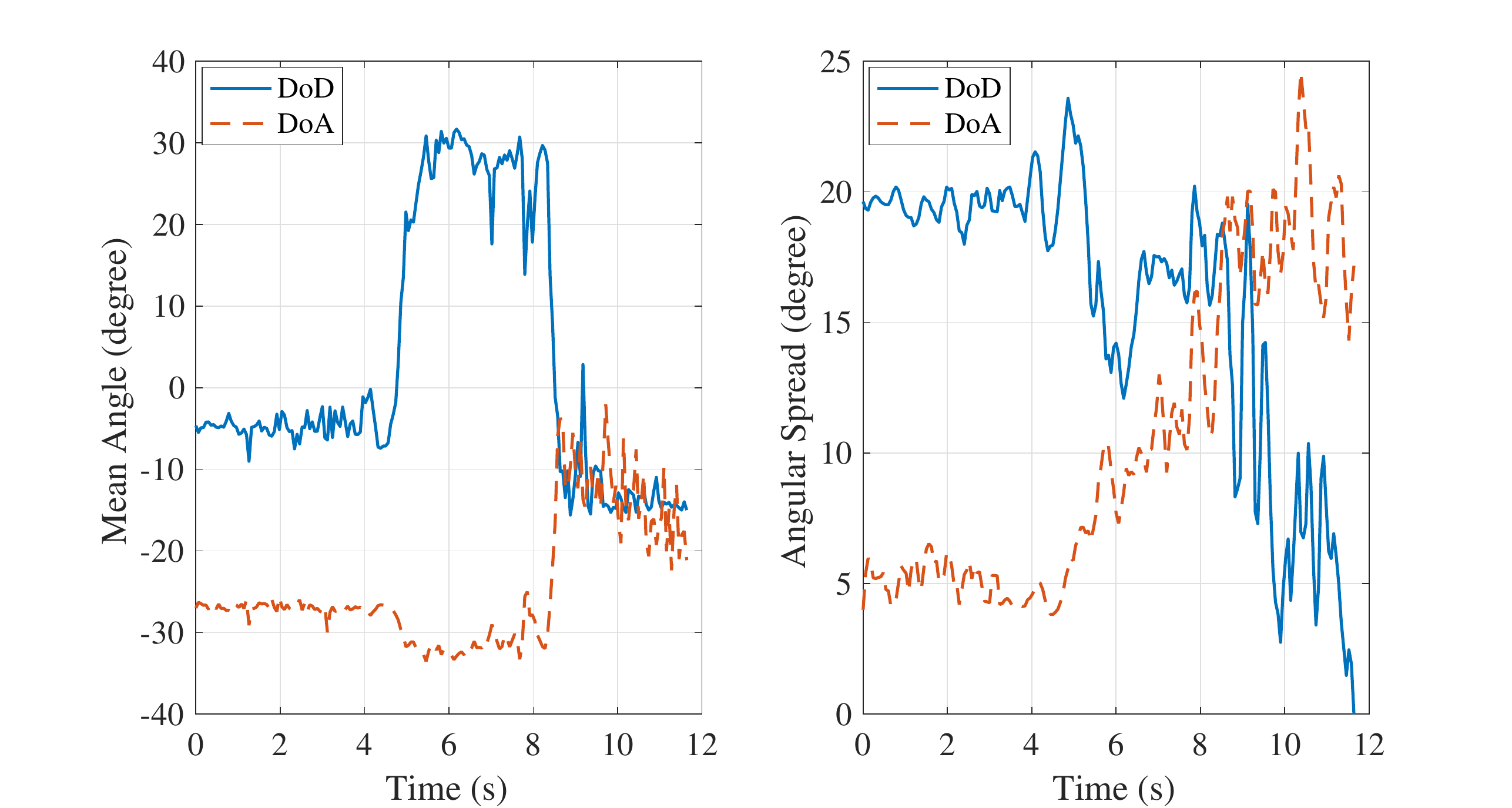}\caption{Mean angles and angular spreads vs time}\label{fig:angstats}
\end{figure}

\subsection{Case 1 : Blocking Objects}

In this scenario, we observe two main MPCs while the channel is idle; the LOS path and a reflection as shown in Figure \ref{fig:dynamic_1}. The actions during this measurement can be listed as;
\begin{itemize}
  \item The measurements start with an idle channel,
  \item a bus enters to the street moving right to left,
  \item the bus blocks the LOS at $t=5s$,
  \item the bus blocks both the LOS and the reflection at $t=8.5s$,
  \item the bus stops in front of the RX while still blocking the two main paths.
\end{itemize}

Figure \ref{fig:paths_1} shows the PDP versus time, at $t=0s$ the two dominant paths, LOS and the reflection can be seen at the delays of \SI{51}{m} and \SI{57.75}{m} respectively. The corresponding angular spectra for TX and RX are shown in Figure \ref{fig:angles_1}. The LOS path is at $[\theta_{TX},\theta_{RX}]=[-15,-25]$  and reflection, is at $[\theta_{TX},\theta_{RX}]=[35,-35]$. 

\begin{figure}[tbp]
        \centering\includegraphics[width=1\linewidth]{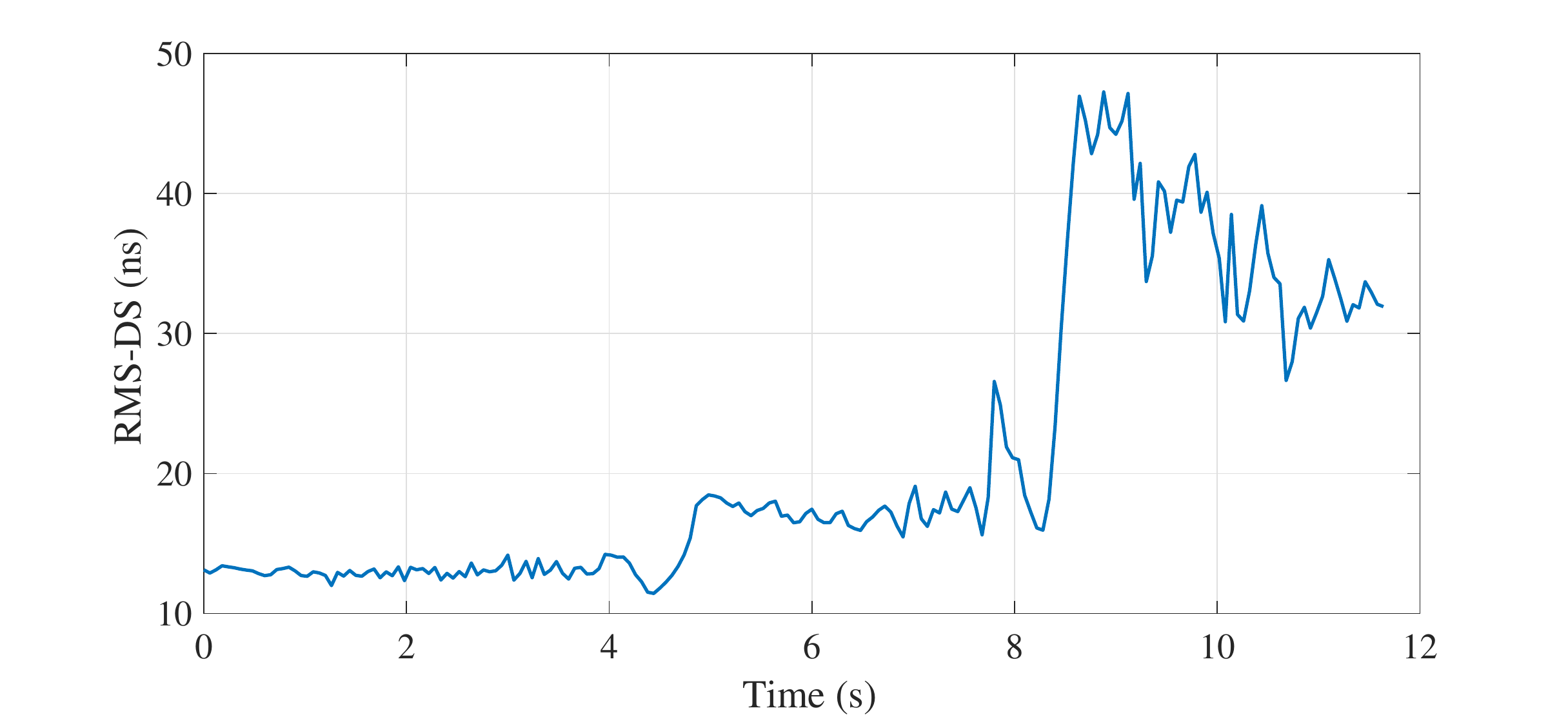}\caption{Root mean square delay spread vs time}\label{fig:rms_ds}
\end{figure}

The effects of a bus blocking the RX on the delay and angular dispersion are visible in the figures as well. As seen in Figure \ref{fig:angstats}, due to the two dominant paths with similar DoAs and different DoDs, initially the angular spread for DoA is as small as $5^\circ$ while it is $20^\circ$ for DoD. Once the bus blocks the RX sector almost completely, the only strong paths are through the bus or diffracted around the bus. Consequently, The angular spread for the ådirection of departure (DoD) is limited compared to the direction of arrival (DoA). Once both dominant paths are blocked the root mean square delay spread (RMS-DS) increases from \SI{12}{ns} to approximately \SI{40}{ns}, Figure \ref{fig:rms_ds}.

\begin{figure}[tbp]
        \centering\includegraphics[width=1\linewidth]{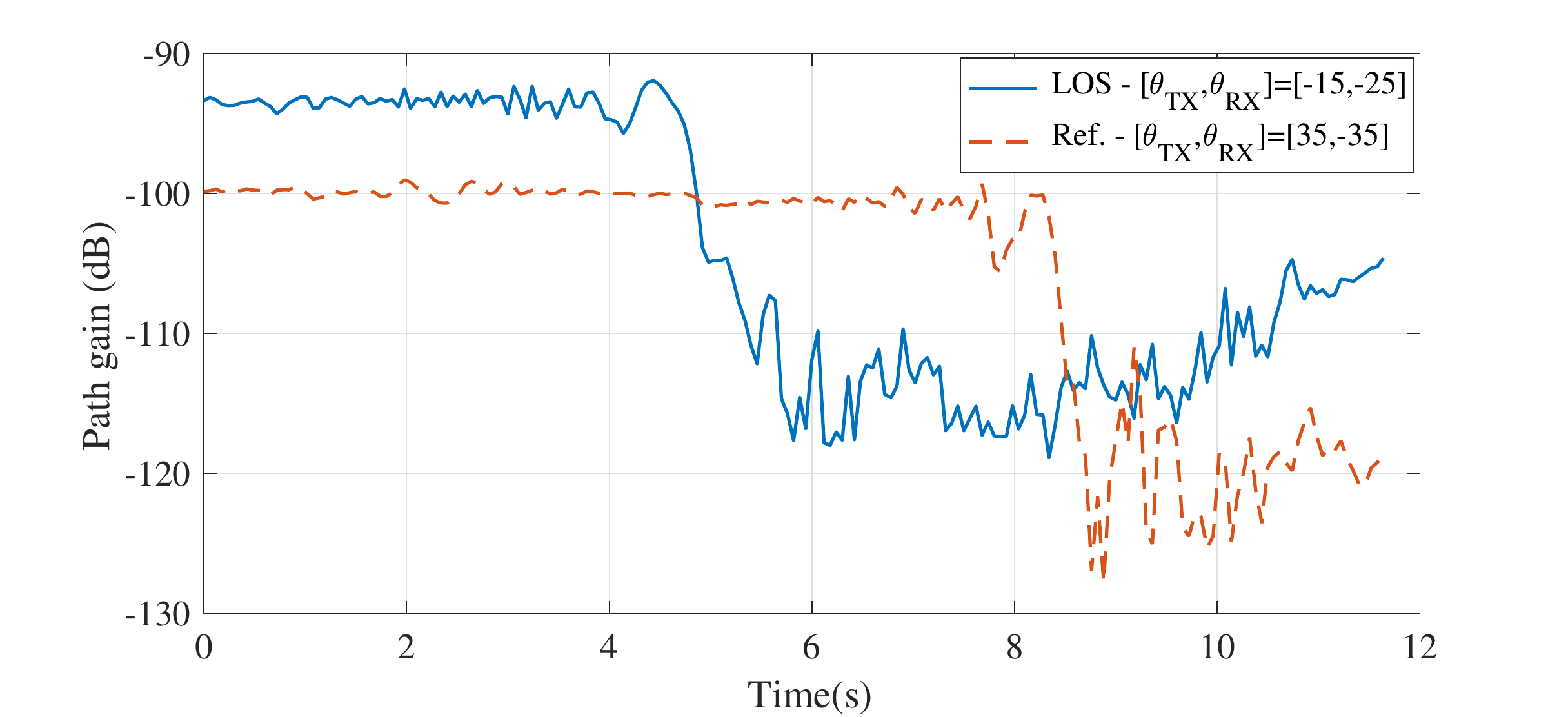}\caption{Path gains for best two beam-pairs vs time}\label{fig:beam_gains}
\end{figure}

Figure \ref{fig:beam_gains} shows the temporal variations of the path gains for the two dominant RX-TX beam pairs. Between $t=5s$ and $t=8.5s$, if both TX and RX can select the best beam then the excess loss is \SI{9}{dB} compared to the \SI{20}{dB} if we use fixed beams. This shows that beam direction adaptation is essential not only for  mm-wave systems in mobile scenarios, but also with fixed TX and RX if significant blockage can occur.

\subsection{Case 2 : Moving Scatterers}

\begin{figure}[tbp]
        \centering\includegraphics[width=1\linewidth]{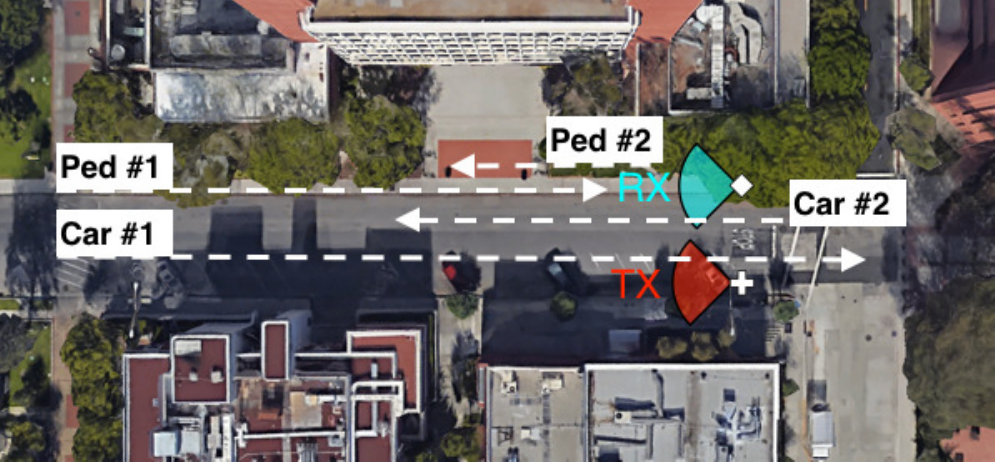}\caption{Moving Scatterers}\label{fig:dynamic2}
\end{figure}

\begin{figure}[tbp]
        \centering\includegraphics[width=0.9\linewidth, viewport=50 185 550 580, clip=true]{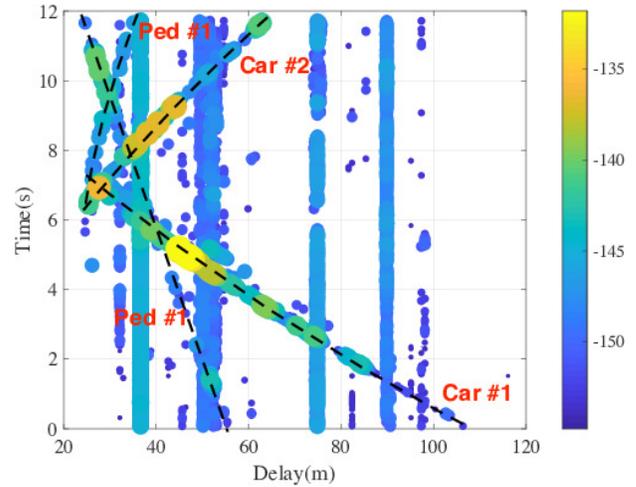}\caption{Temporal evolution of the multipath components}\label{fig:paths}
\end{figure}

\begin{figure}[htbp]
        \centering\includegraphics[width=1.04\linewidth]{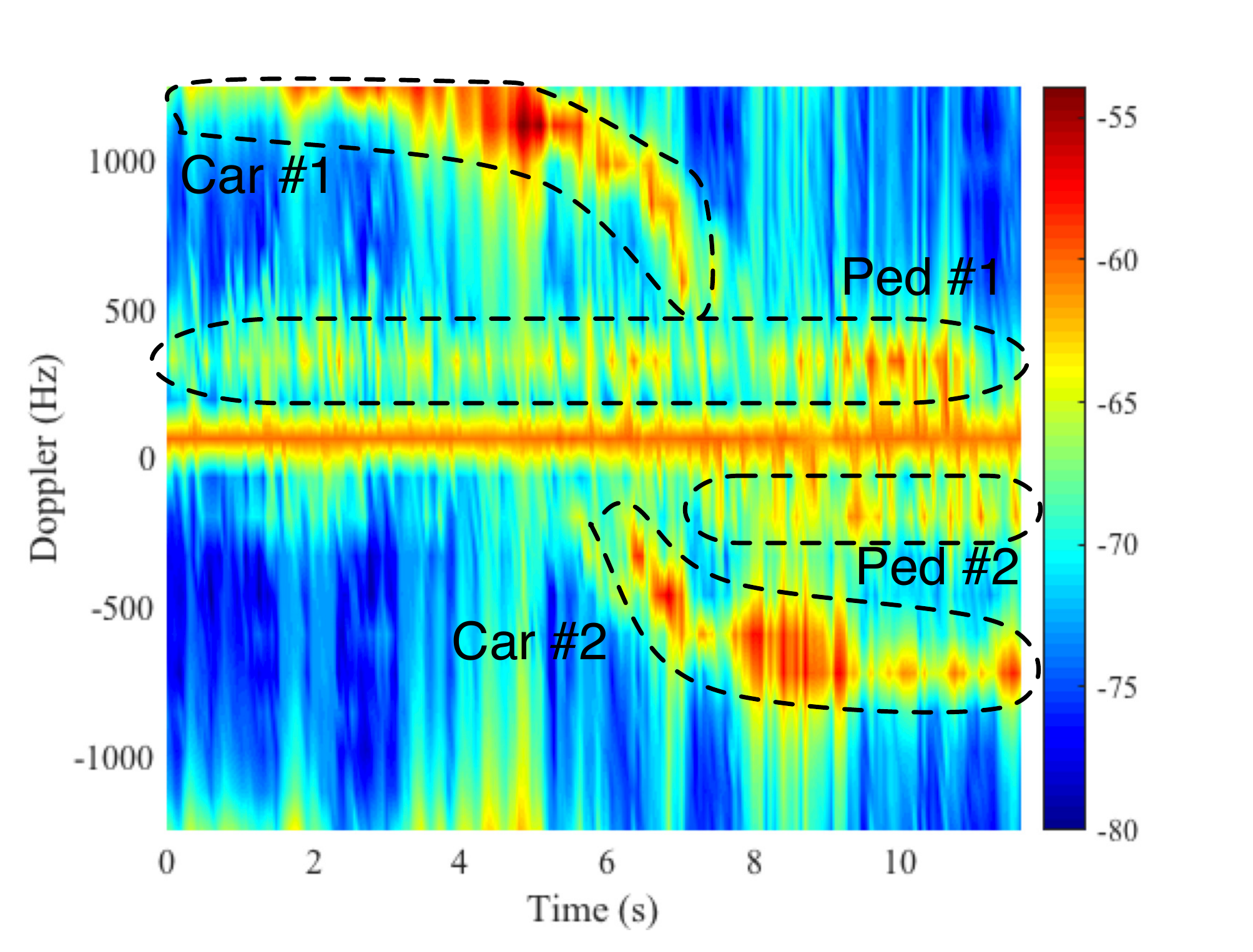}\caption{Doppler spectrum for case 2}\label{fig:doppler}
\end{figure}

In a second measurement, both TX and RX are placed facing the same direction so that they are out of each other's visible azimuth range. We can thus observe moving scatterers without a dominant LOS. Figure \ref{fig:dynamic2} shows the details about the measurement snapshot presented in this section. During \SI{12}{s}, we observe 4 moving objects; Car \#1 and Pedestrian \#1 are coming towards the TX and RX starting from $t=0s$. The tracked MPCs corresponding these two objects are marked as Car \#1 and Ped \#1 in Figure \ref{fig:paths}. Furthermore, these objects are also clearly visible in the Doppler spectrum with positive Doppler shifts as expected, see Figure \ref{fig:doppler}. 

After $t=6s$, Car \#2 and Ped \#2 enter the picture and move away from TX and RX. 
Between time $t=9s$ and $t=11s$, Ped \#1 blocks the path from Car \#2 to RX. Their effect on the Doppler spectrum is also marked in Figure \ref{fig:doppler}. The speeds of all four objects estimated speeds from the Doppler spectrum matches with the video recordings captured simultaneously with the measurements.

\begin{figure}[htbp]
        \centering\includegraphics[width=1\linewidth]{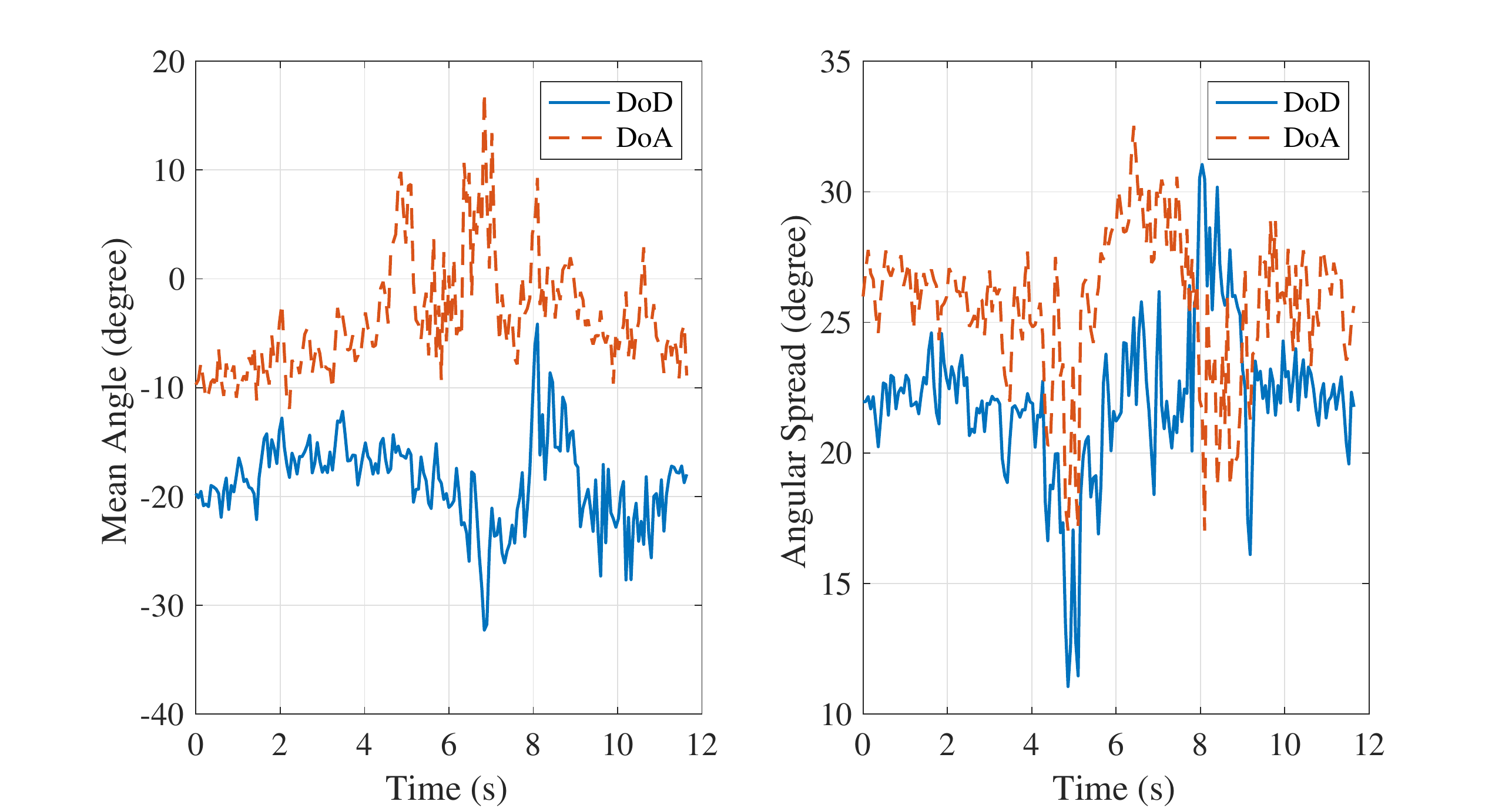}\caption{Mean angles and angular spreads vs time for Case 2}\label{fig:case2_ang_stats}
\end{figure}

\begin{figure}[htbp]
        \centering\includegraphics[width=1\linewidth]{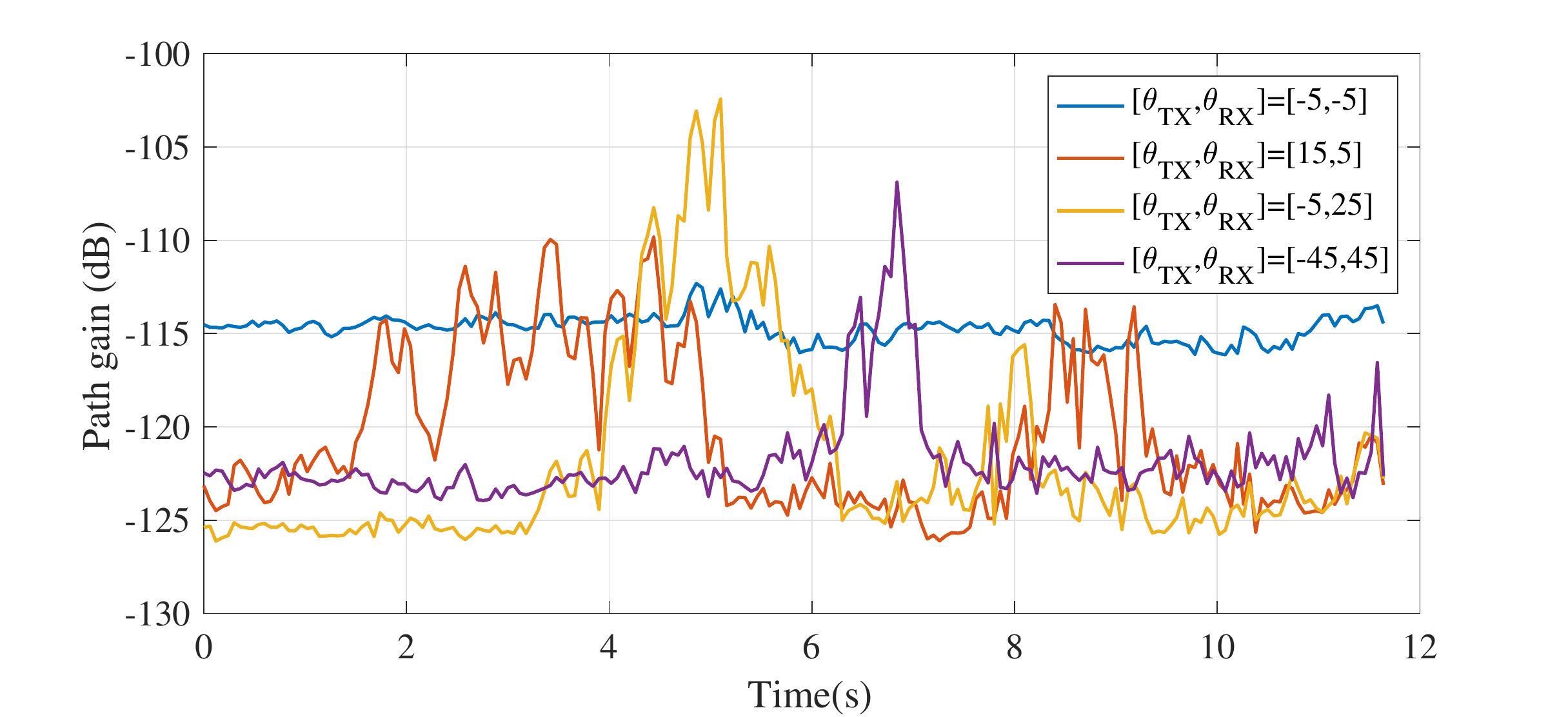}\caption{Path gains for best beam-pairs vs  time for Case 2}\label{fig:beam_gains_case2}
\end{figure}

Due to fast moving objects in the environment, and their interactions with each other (e.g., Ped \#1 shadowing Car \#2 ) the angular spectra for the TX and the RX change quickly. As seen in Figure \ref{fig:case2_ang_stats}, between $t=5$ and $t=8$, the mean DoA varies in between $-10^\circ$  and $10^\circ$ continuously. During the same time period, angular spread for DoD varies from $12^\circ$ to $31^\circ$.  Figure \ref{fig:beam_gains_case2} shows the evolution of the path gains for four different beam pairs over time. The beam pair \#1 at $[\theta_{TX},\theta_{RX}]=[-5,5]$ point towards a static reflector and has a roughly constant gain of \SI{-115}{dB} throughout this snapshot. Although at times due to moving reflectors, the other three beams surpass the beam pair  \#1 by as much as \SI{10}{dB}, this improvement is temporary and never lasts more than \SI{1}{s}. Depending on the overhead of finding the best beam and the required time to update both RX and TX beams, it might not be optimum to use an approach which constantly looks for the best beam. Other approaches such as remaining at a beam pair which is ``good''  enough or implementing a trigger hysteresis for beam update should be considered as well.

\subsection{Case 3 : Blocked LOS}

These measurements were performed with TX\_2 and RX\_3 in Figure \ref{fig:loc_all}. In this case, a strong LOS is present with no other significant reflections. During the measurements, the LOS was blocked by a moving bus in between $t=6s$ and $t=8s$. 

\begin{figure}[htbp]
        \centering\includegraphics[width=1\linewidth]{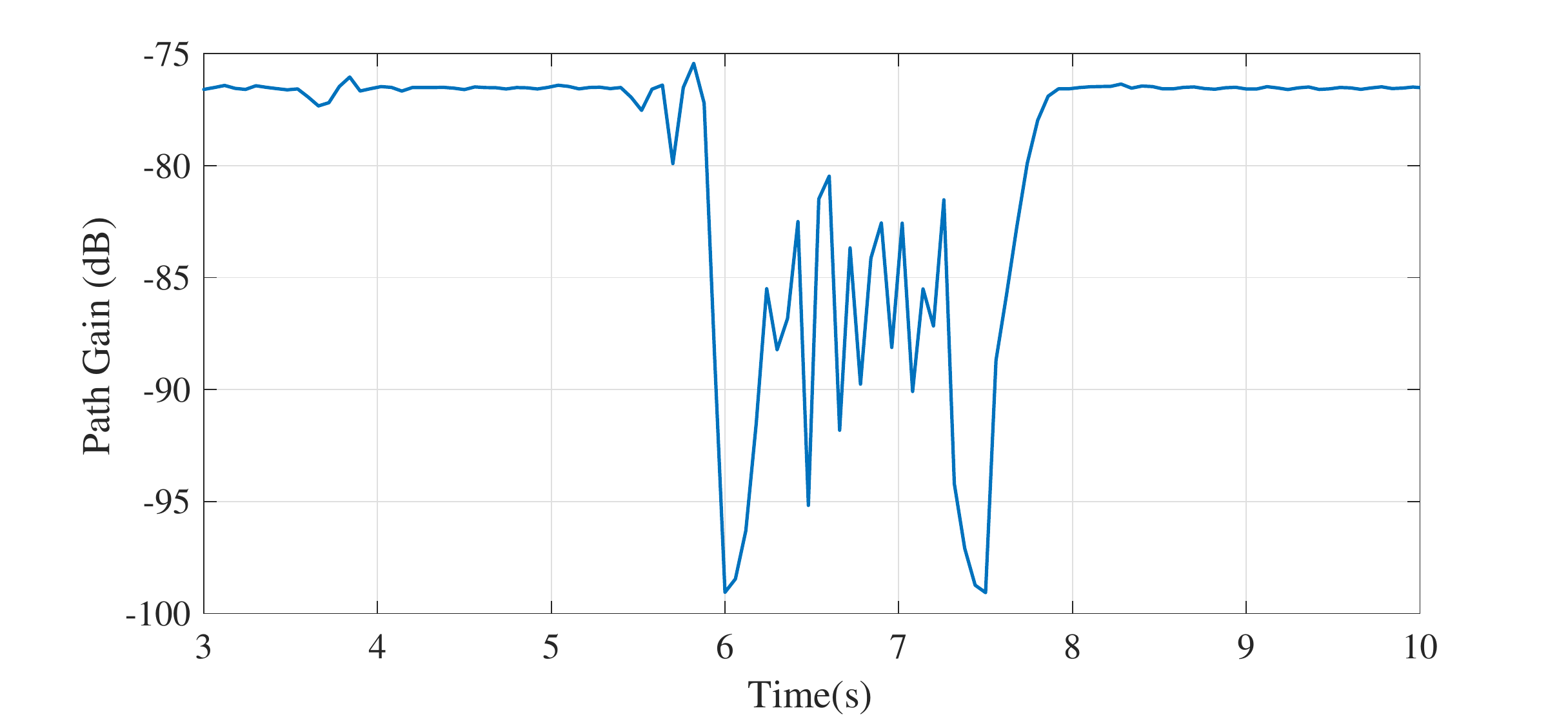}\caption{Path gain vs time for Case 3}\label{fig:case3_pg}
\end{figure}

Figure \ref{fig:case3_pg} shows the time-variant path gain. Due to the metallic parts of the bus in the front and the back, we observe excess losses up to \SI{24}{dB}. In this particular case, it takes \SI{120}{ms} for the excess loss to reach to \SI{24}{dB} and it stays in the shadowing dip for \SI{240}{ms}. As the bus continues to move, the type of material blocking the direct path changes between metal and glass, hence the excess loss varies between \SI{5}{dB} and \SI{15}{dB} for \SI{1.14}{s}. As the back end of the bus blocks the direct path, we observe another fading deep lasting \SI{270}{ms}. Afterwards, the excess loss returns to \SI{0}{dB} in \SI{300}{ms} as the bus clears the area.

\begin{figure}[htbp]
        \centering\includegraphics[width=1\linewidth]{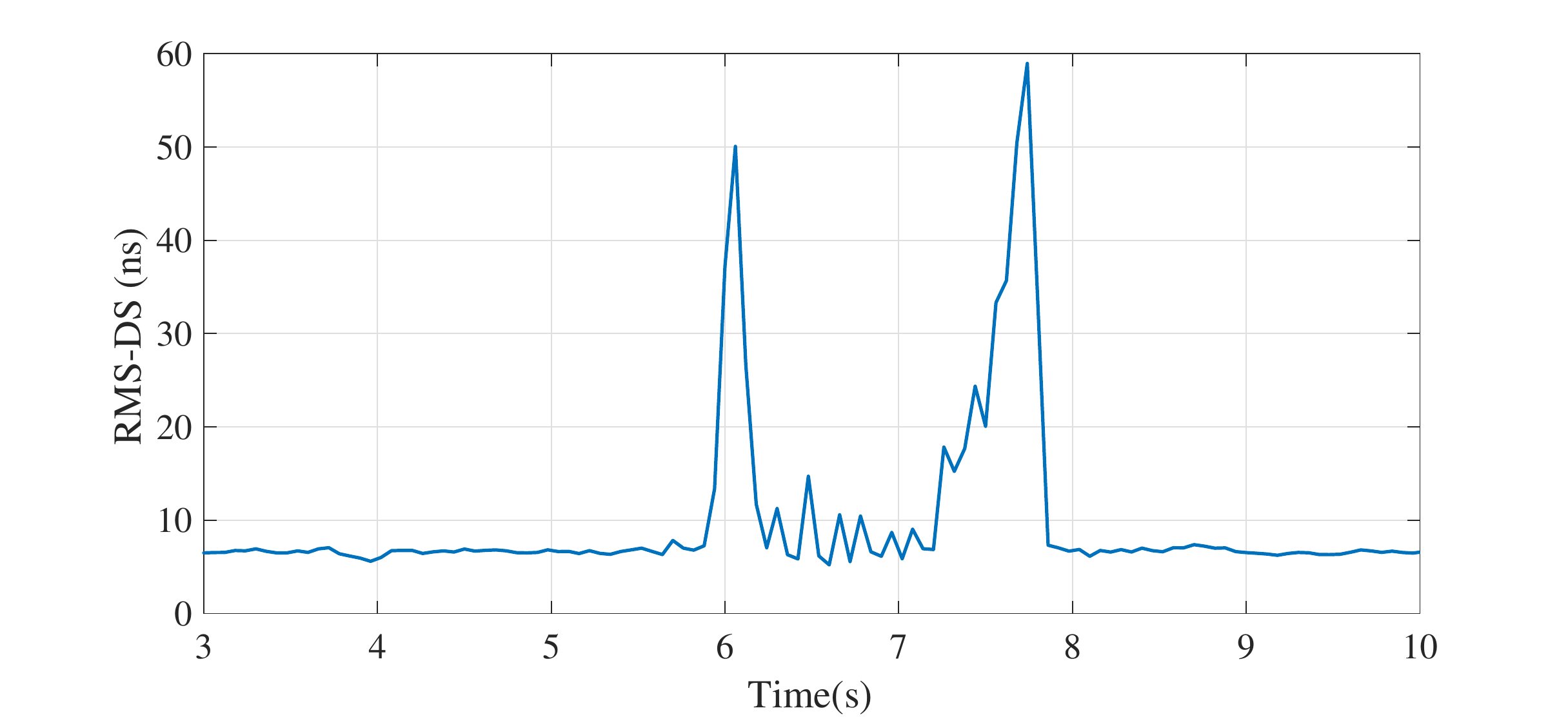}\caption{RMS-DS vs time for Case 3}\label{fig:case3_rms_ds}
\end{figure}

\begin{figure}[htbp]
        \centering\includegraphics[width=1\linewidth]{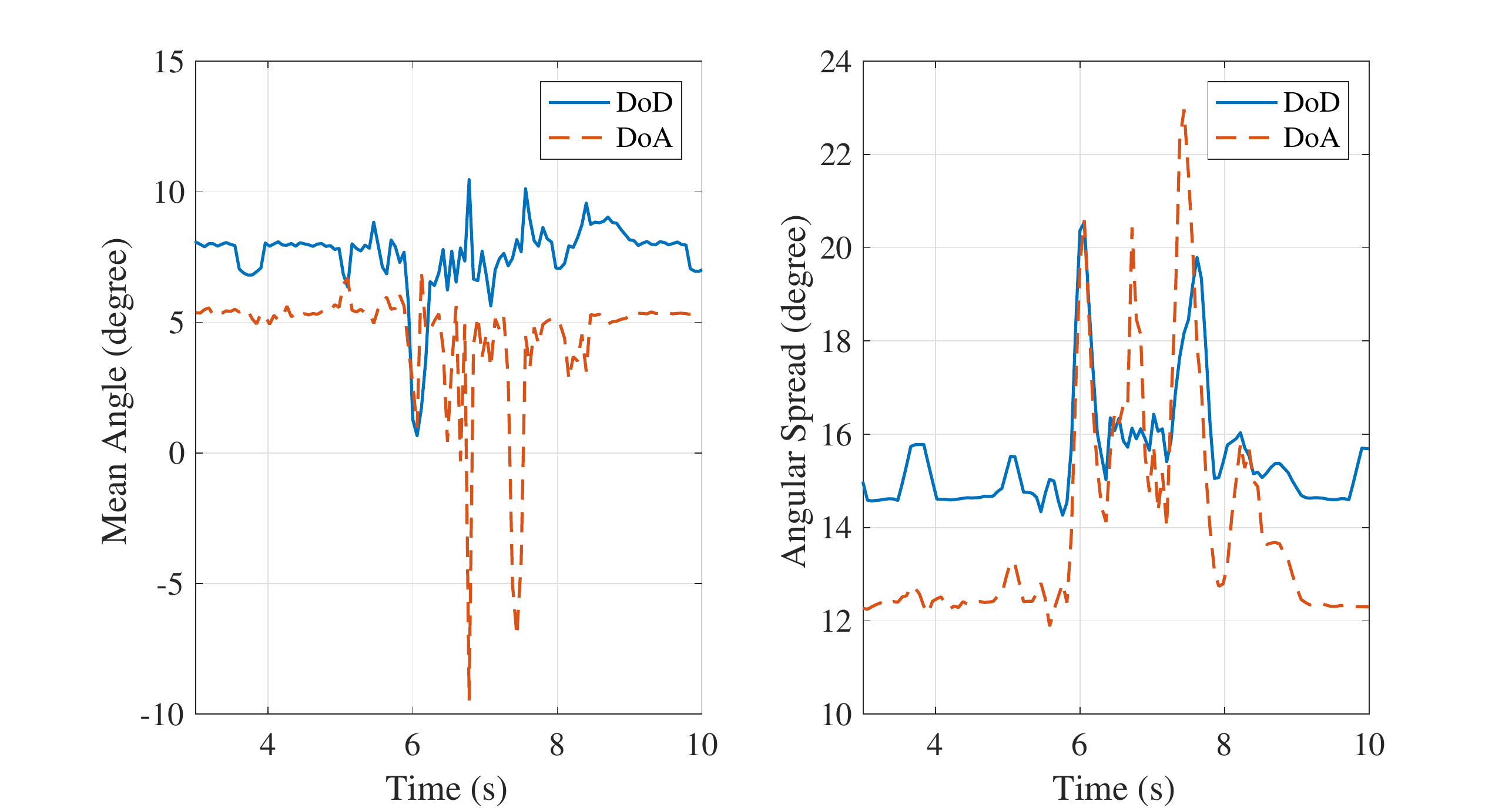}\caption{Mean angles and angular spreads vs time for Case 3}\label{fig:case3_ang_stats}
\end{figure}

Figure \ref{fig:case3_rms_ds} shows the time-varying RMS-DS for the same duration. While the channel is idle the RMS-DS is around \SI{6.5}{ns}. The RMS-DS is increasing as high as \SI{50}{ns} during the aforementioned fading dips observed as the bus enters and exits the area between the TX and RX. During the rest of the blockage, the RMS-DS varies \SI{6}{ns} to \SI{15}{ns}. Figure \ref{fig:case3_ang_stats} shows the angular statistics. Similar to the RMS-DS, the effect of the moving bus on the angular spectra is prominent while the bus enters or exits the visibility of the TX and the RX. Additionally, as the bus is in view (i.e. between $t=6s$ and $t=8s$) both the mean angles and the angular spread vary continuously.

\section{Conclusions}\label{sec_conc}

In this paper, we presented results from a double-directionally resolved measurement campaign at mm-wave frequencies for an outdoor microcellular scenario. The unique capabilities of the channel sounder allowed phase-coherent measurements enabling investigations into the temporal channel characteristics. We provided results for three scenarios focusing on angular spectrum, delay spread, and Doppler spectrums. We also gave examples of evolutions of these channel parameters in a dynamic channel. 

\section*{Acknowledgements}
Part of this work was supported by grants from the National Science Foundation and National Institute of Standards and Technology. The authors would like to thank Dimitris Psychoudakis, Thomas Henige, Robert Monroe for their contribution in the development of the channel sounder.

\bibliographystyle{IEEEtran}
\bibliography{mmwave,dynamic}

\begin{thebibliography}{10}
\providecommand{\url}[1]{#1}
\csname url@samestyle\endcsname
\providecommand{\newblock}{\relax}
\providecommand{\bibinfo}[2]{#2}
\providecommand{\BIBentrySTDinterwordspacing}{\spaceskip=0pt\relax}
\providecommand{\BIBentryALTinterwordstretchfactor}{4}
\providecommand{\BIBentryALTinterwordspacing}{\spaceskip=\fontdimen2\font plus
\BIBentryALTinterwordstretchfactor\fontdimen3\font minus
  \fontdimen4\font\relax}
\providecommand{\BIBforeignlanguage}[2]{{%
\expandafter\ifx\csname l@#1\endcsname\relax
\typeout{** WARNING: IEEEtran.bst: No hyphenation pattern has been}%
\typeout{** loaded for the language `#1'. Using the pattern for}%
\typeout{** the default language instead.}%
\else
\language=\csname l@#1\endcsname
\fi
#2}}
\providecommand{\BIBdecl}{\relax}
\BIBdecl

\bibitem{forecast2017cisco}
C.~V. Forecast, ``Cisco visual networking index: Global mobile data traffic
  forecast update, 2016--2021 white paper,'' \emph{Cisco Public Information},
  2017.

\bibitem{Molisch_2016_eucap}
A.~F. Molisch, A.~Karttunen, R.~Wang, C.~U. Bas, S.~Hur, J.~Park, and J.~Zhang,
  ``Millimeter-wave channels in urban environments,'' in \emph{2016 10th
  European Conference on Antennas and Propagation (EuCAP)}, April 2016, pp.
  1--5.

\bibitem{Roh2014millimeter}
W.~Roh, J.~Y. Seol, J.~Park, B.~Lee, J.~Lee, Y.~Kim, J.~Cho, K.~Cheun, and
  F.~Aryanfar, ``Millimeter-wave beamforming as an enabling technology for 5g
  cellular communications: theoretical feasibility and prototype results,''
  \emph{IEEE Communications Magazine}, vol.~52, no.~2, pp. 106--113, February
  2014.

\bibitem{bas_realjournal_2017}
C.~U. Bas and et.al., ``{A Real-Time Millimeter-Wave Phased Array MIMO Channel
  Sounder for Dynamic Measurements },'' to be submitted.

\bibitem{bas_2017_realtime}
C.~U. Bas, R.~Wang, D.~Psychoudakis, T.~Henige, R.~Monroe, J.~Park, J.~Zhang,
  and A.~F. Molisch, ``{A Real-Time Millimeter-Wave Phased Array MIMO Channel
  Sounder},'' in \emph{Vehicular Technology Conference, 2017. VTC 2017-Fall.
  IEEE}, September 2017.

\bibitem{MacCartney_2017_flexible}
G.~R. MacCartney and T.~S. Rappaport, ``A flexible millimeter-wave channel
  sounder with absolute timing,'' \emph{IEEE Journal on Selected Areas in
  Communications}, vol.~PP, no.~99, pp. 1--1, 2017.

\bibitem{hur_synchronous_2014}
S.~Hur, Y.~J. Cho, J.~Lee, N.-G. Kang, J.~Park, and H.~Benn, ``{Synchronous
  channel sounder using horn antenna and indoor measurements on 28 GHz},'' in
  \emph{2014 IEEE International Black Sea Conference on Communications and
  Networking (BlackSeaCom)}, May 2014, pp. 83--87.

\bibitem{Haneda_2016_omni}
K.~Haneda, S.~L.~H. Nguyen, J.~JŠrvelŠinen, and J.~Putkonen, ``Estimating the
  omni-directional pathloss from directional channel sounding,'' in \emph{2016
  10th European Conference on Antennas and Propagation (EuCAP)}, April 2016,
  pp. 1--5.

\bibitem{Marinier1998temporal}
P.~Marinier, G.~Y. Delisle, and C.~L. Despins, ``Temporal variations of the
  indoor wireless millimeter-wave channel,'' \emph{IEEE Transactions on
  Antennas and Propagation}, vol.~46, no.~6, pp. 928--934, Jun 1998.

\bibitem{maccartney2017rapid}
G.~R. MacCartney~Jr, T.~S. Rappaport, and S.~Rangan, ``Rapid fading due to
  human blockage in pedestrian crowds at 5g millimeter-wave frequencies,''
  \emph{arXiv preprint arXiv:1709.05883}, 2017.

\bibitem{collonge2004influence}
S.~Collonge, G.~Zaharia, and G.~E. Zein, ``Influence of the human activity on
  wide-band characteristics of the 60 ghz indoor radio channel,'' \emph{IEEE
  Transactions on Wireless Communications}, vol.~3, no.~6, pp. 2396--2406,
  2004.

\bibitem{Park_2017_vehicular}
J.-J. Park, J.~Lee, J.~Liang, K.-W. Kim, K.-C. Lee, and M.-D. Kim,
  ``{Millimeter Wave Vehicular Blockage Characteristics Based on 28 GHz
  Measurements},'' in \emph{Vehicular Technology Conference, 2017. VTC
  2017-Fall. IEEE}, September 2017.

\bibitem{Weiler_et_al_2016_quasideterministic}
R.~J. Weiler, M.~Peter, W.~Keusgen, A.~Maltsev, I.~Karls, A.~Pudeyev,
  I.~Bolotin, I.~Siaud, and A.-M. Ulmer-Moll, ``Quasi-deterministic
  millimeter-wave channel models in miweba,'' \emph{EURASIP Journal on Wireless
  Communications and Networking}, vol. 2016, no.~1, p.~84, 2016.

\bibitem{Weiler_et_al_2016_WCL}
R.~J. Weiler, M.~Peter, W.~Keusgen, K.~Sakaguchi, and F.~Undi, ``Environment
  induced shadowing of urban millimeter-wave access links,'' \emph{IEEE
  Wireless Communications Letters}, vol.~5, no.~4, pp. 440--443, Aug 2016.

\bibitem{Semkin_et_al_2015_EuCAP}
V.~Semkin, U.~Virk, A.~Karttunen, K.~Haneda, and A.~V. RŠisŠnen, ``E-band
  propagation channel measurements in an urban street canyon,'' in \emph{2015
  9th European Conference on Antennas and Propagation (EuCAP)}, May 2015, pp.
  1--4.

\bibitem{Sato_et_al_2001_V2I_RoF}
K.~Sato, M.~Fujise, R.~Tachita, E.~Hase, and T.~Nose, ``Propagation in rof
  road-vehicle communication system using millimeter wave,'' in \emph{IVEC2001.
  Proceedings of the IEEE International Vehicle Electronics Conference 2001.
  IVEC 2001 (Cat. No.01EX522)}, 2001, pp. 131--135.

\bibitem{Friese1997multitone}
M.~Friese, ``Multitone signals with low crest factor,'' \emph{Communications,
  IEEE Transactions on}, vol.~45, no.~10, pp. 1338--1344, Oct 1997.

\bibitem{bas_2017_microcell}
C.~U. Bas, R.~Wang, S.~Sangodoyin, S.~Hur, K.~Whang, J.~Park, J.~Zhang, and
  A.~F. Molisch, ``{28 GHz Microcell Measurement Campaign for Residential
  Environment },'' in \emph{{2017 IEEE Global Communications Conference
  (GLOBECOM)}}, December 2017.

\bibitem{wang_2017_stationarity}
R.~Wang, C.~U. Bas, S.~Sangodoyin, S.~Hur, K.~Whang, J.~Park, J.~Zhang, and
  A.~F. Molisch, ``{Stationarity Region of Mm-Wave Channel Based on Outdoor
  Microcellular Measurements at 28 GHz},'' in \emph{{Military Communications
  Conference, MILCOM 2017- 2017 IEEE}}, October 2017.

\end{thebibliography}

\end{document}